\documentclass[twoside]{article}
\usepackage{graphicx}
\usepackage{fancyhdr}
\usepackage{multicol}
\usepackage{styl-ap}

\begin{document}
\title{Computational Schemes for the Propagation of \\
Ultra High Energy Cosmic Rays}
\author{Roberto Aloisio\work{1,2}}
\workplace{INAF, Osservatorio Astrofisico Arcetri, I-50125 Arcetri (Firenze) Italy
\next
INFN, Laboratori Nazionali del Gran Sasso, I-67010 Assergi (L'Aquila) Italy}
\mainauthor{aloisio@arcetri.astro.it}
\maketitle

\begin{abstract}
We discuss the problem of ultra high energy particles propagation
in astrophysical backgrounds. We present two different computational schemes based 
on both kinetic and Monte Carlo approaches. The kinetic approach is an analytical
computation scheme based on the hypothesis of continuos energy losses while 
the Monte Carlo scheme takes into account also the stochastic nature of particle
interactions. These schemes, that give quite reliable results, enable the computation 
of fluxes keeping track of the different primary and secondary components, providing 
a fast and useful workbench to study Ultra High Energy Cosmic Rays. 
\end{abstract}


\begin{multicols}{2}

\section{Introduction and Conclusions}
Ultra High Energy Cosmic Rays (UHECR) are the most energetic particles observed in nature, with energies up to 
few$\times 10^{20}$ eV. Nowadays the experimental study of UHECR 
is conducted by three different experiments: Auger in Argentina, HiRes and Telescope Array in the USA. 

The propagation of UHECR from the source to the observer is conditioned 
by their interactions with astrophysical backgrounds: the Cosmic Microwave
Background (CMB) and the Extragalactic Background Light (EBL). Understanding the key features of the propagation is 
of paramount importance to interpret experimental observations paving the way for the discovery of the astrophysical origin 
of these fascinating particles. 

Several features of the observed spectrum can be directly linked to the chemical composition of UHECR and to their
sources (Greisen 1966, Zatsepin \& Kuzmin 1966, Berezinsky et al. 2006a, Aloisio et. al. 2007, Aloisio \& Boncioli 2011). 
Among such features particularly important is the Greisin, Zatsepin and Kuzmin (GZK) 
suppression of the flux, an abrupt depletion of the observed proton spectrum, arising at energies 
$E\simeq 5\times 10^{19}$ eV, due to the interaction of UHE protons with the CMB radiation field (Greisen 1966, 
Zatsepin \& Kuzmin 1966). The GZK suppression, as follows from the 
original papers, is referred to protons and it is due to the photo-pion production process on the CMB radiation field 
$(p+\gamma_{CMB}\to\pi+p)$. In the case of nuclei the expected flux also shows a suppression at the highest energies 
that, depending on the nuclei specie, is due to the photo-disintegration process on the CMB and EBL radiation fields 
$(A + \gamma_{CMB,EBL} \to (A - nN) + nN)$ (Aloisio et. al. 2012a). 
Another important feature in the spectrum that can be directly linked with the nature of the primary particles and their 
origin (galactic/extra-galactic) is the pair-production dip (Berezinsky et al. 2006a, Aloisio et al. 2007). This feature is present 
only in the spectrum of UHE extragalactic protons and, as the GZK, is a direct consequence of the proton interaction with 
the CMB radiation field, in particular the dip brings a direct imprint of the pair production process 
$p + \gamma_{CMB} \to p + e^{+} + e^{-}$ suffered by protons.

From the experimental point of view the situation is far from being clear with different experiments claiming
contradictory results. The HiRes experiment, not anymore taking data, showed a proton dominated spectrum till 
the highest energies (HiRes collaboration 2008, 2010) while the Auger observations show an heavy mass composition 
at energies $E>4\times 10^{18}$ eV (Auger collaboration 2010). 

This puzzling situation, with different experiments favoring different scenarios, shows once more the importance 
of a systematic study of UHECR propagation in astrophysical backgrounds. In the present paper we 
will review the main points of two alternative computation schemes which enable the determination of the 
fluxes expected on earth fixing the injection spectrum and the distribution of sources. These two schemes 
are based on different approaches to model particle interactions with backgrounds: the continuum 
energy losses (CEL) approximation, which is the base for the kinetic approach, and the Monte Carlo (MC) 
technique. 

As we will discuss in the following these two different schemes give reliable results that, in the framework 
of the different assumptions, agree each other offering a suitable theoretical framework to study
experimental results unveiling the intimate nature of UHECR.

\section{Kinetic Equations} 
The main assumption under which the kinetic theory is build is the 
CEL approximation (Berezinsky et al 1990), through which particle interactions are 
treated as a continuum process that continuously depletes the particles energy. 

UHECR propagating through astrophysical backgrounds suffer 
different interaction processes 

\begin{itemize}
\item{{\it protons}} - UHE protons interact only with the CMB radiation field giving rise to the two processes of pair
production and photo-pion production. Both of these reactions can be treated in the CEL hypothesis. 

\item{{\it nuclei}} - UHE nuclei interact with the CMB and EBL radiation fields, suffering the process of pair
production, for which only CMB is relevant, and photo-disintegration, that involves both backgrounds CMB 
and EBL. While the first process can be treated in the CEL hypothesis, being the 
nuclei specie conserved, the second cannot be, producing a change in the nucleus specie. Following Aloisio et. al. 2012a, 
in the framework of the kinetic approach, we will treat the photo-disintegration process as a "decaying" process 
that simply depletes the flux of the propagating nucleus. 
\end{itemize}
 
Taking into account all energy losses processes we can describe the propagation of protons and nuclei through
kinetic equations of the type: 
\begin{equation}
\frac{\partial n_p(\Gamma,t)}{\partial t} - \frac{\partial}{\partial 
\Gamma} \left [ b_p(\Gamma,t)n_p(\Gamma,t) \right ] = Q_p(\Gamma,t)
\label{eq:kin_p}
\end{equation}   
\begin{equation}
\frac{\partial n_{A}(\Gamma,t)}{\partial t} - \frac{\partial}{\partial \Gamma}
\left [ n_{A}(\Gamma,t) b_{A}(\Gamma,t) \right ] + \frac{n_{A}(\Gamma,t)}
{\tau_{A}(\Gamma,t)}  = Q_{A}(\Gamma,t)
\label{eq:kin_A}
\end{equation}
where $n$ is the equilibrium distribution of particles, $b$ are the energy losses (adiabatic expansion of the Universe
and pair/photo-pion production for protons or only pair-production for nuclei) $Q$ is the injection of freshly
accelerated particles and, in the case of nuclei, also the injection of secondary particles produced by 
photo-disintegration (see below). 

The energy losses $b$ for protons or nuclei depend only on the CMB field and in the CEL hypothesis can be
computed  analytically (Berezinsky et. al. 2006a, Aloisio et. al . 2007, Aloisio et. al. 2012a).

The second process that affects nuclei propagation is photo-disintegration over CMB and EBL backgrounds. This process
is treated as a decaying process that depletes the flux of nuclei, it enters in the kinetic equation (see equation 
(\ref{eq:kin_A})) through a sort of "life-time" of the nucleus under the photo-disintegration process. This 
"life-time" corresponds to the mean time needed to a nucleus of Lorentz factor $\Gamma$ and atomic mass number 
$A$ to lose, at least, one of its nucleons: 
\begin{equation}
\frac{1}{\tau_A}=\frac{c}{2\Gamma^2}
\int_{\epsilon_0(A)}^{\infty} d\epsilon_r \sigma(\epsilon_r,A)\nu(\epsilon_r)\epsilon_r
\int_{\epsilon_r/(2\Gamma)}^{\infty} d\epsilon \frac{n_{bkg}(\epsilon)}{\epsilon^2}
\label{eq:loss2}
\end{equation}
where $\sigma(\epsilon_r,A)$ is the photo-disintegration cross-section and $\nu(\epsilon_r)$ is 
the molteplicity associated with this process, namely the average number of nucleons extracted 
from the nucleus by a single interaction and $n_{bkg}=n_{CMB}+n_{EBL}$. 
The dependence on red-shift of $\tau_A$ directly follows from the evolution with red-shift of the background 
photon densities $n_{CMB}$ and $n_{EBL}$. In the case of CMB this dependence is known analytically 
while for the EBL one should refer to evolution models (in our computations we have used the model 
by Stecker et. al. 2006).

One important feature of the photo-disintegration process is that it starts to contribute to the propagation 
of nuclei at a Lorentz factor that is almost independent of the nuclei specie $\Gamma_{cr}\simeq 2\times 10^{9}$
(Aloisio et al. 2012a). This is an important general characteristic of nuclei photo-disintegration process from which 
we can immediately deduce the dependence on the nucleus specie of the energy corresponding to the 
photo-disintegration suppression of the flux: 
$E^A_{cut} = A m_N \Gamma_{cr}$
being $A$ the atomic mass number of the nucleus and $m_N$ the proton mass. From this expression for $E^A_{cut}$ 
it is evident how the flux behavior could bring informations on the chemical composition of the UHECR,
in the case of Helium ($A=4$) the suppression is expected around energies $E \simeq 10^{19}$ eV 
while in the case of Iron ($A=56$) the suppression is expected at higher energies $E\simeq 10^{20}$ eV. 

Let us discuss now the generation function $Q_A(\Gamma,t)$ in the right hand side
of Eq.~(\ref{eq:kin_A}). One should distinguish among primary nuclei, i.e. nuclei 
accelerated at the source and injected in the intergalactic space, and secondary 
nuclei and nucleons, i.e. particles produced as secondaries in the photo-disintegration
chain.  In the case of primaries the injection function is an assumption of the source 
model, while the injection of secondaries should be modeled 
taking into account the characteristics of the photo-disintegration process. 
The dominant process of photo-disintegration is the one nucleon ($N$)
emission, namely the process $(A+1) +\gamma_{bkg} \to A+N$, this follows directly 
from the behavior of the photo-disintegration cross-section (see Aloisio et al. 2012a and 
references therein) that shows the giant dipole resonance corresponding 
to one nucleon emission. Moreover, at the typical energies of UHECR ($E>10^{17}$ eV) 
one can safely neglect the nucleus recoil so that photo-disintegration will conserve 
the Lorentz factor of the particles. Therefore the production rate of secondary 
$A-$nucleus and $A-$associated nucleons will be given by
\begin{equation}
Q_A(\Gamma,z)= Q_p^A(\Gamma,z)=
\frac{n_{A+1}(\Gamma,z)}{\tau_{A+1}(\Gamma,z)}
\label{eq:injA}
\end{equation}
where $\tau_{A+1}$ is the photo-disintegration life-time of the nucleus father $(A+1)$ 
and $n_{A+1}$ is its equilibrium distribution, solution of the kinetic equation
(\ref{eq:kin_A}). 

Using equation (\ref{eq:injA}) we can build a system of coupled differential equations that
starting from primary injected nuclei $(A_0)$ follows 
the complete photo-disintegration chain for all secondary nuclei $(A<A_0)$ and nucleons. Clearly 
secondary protons\footnote{Neutrons decay very fast into protons, so we will always refer to 
secondary protons.} propagation will be described by the proper kinetic equation (\ref{eq:kin_p}) 
with an injection term given by (\ref{eq:injA}). 
The solution of the kinetic equation for protons and nuclei can be worked out analytically. In the case
of protons:
\begin{equation}
n_p(\Gamma,z)=\int_{z}^{z_{max}} \frac{dz'}{(1+z')H(z')} 
Q_p(\Gamma',z) \frac{d\Gamma'}{d\Gamma}
\label{eq:np-solut}
\end{equation}
being $Q_p$ the injection of primary protons  or secondary
protons (equation (\ref{eq:injA})) and $\Gamma'=\Gamma'(\Gamma,z)$ is the characteristic function of 
the kinetic equation (Aloisio et al. 2012a). In the case of nuclei:
\begin{equation}
n_A(\Gamma,z)=\int_{z}^{z_{max}} \frac{dz'}{(1+z')H(z')} 
Q_A(\Gamma',z)\frac{d\Gamma'}{d\Gamma} e^{-\eta_{A}(\Gamma',z')}.
\label{eq:nA-solut}
\end{equation}
being, again, $Q_A$ the injection of primary or secondary
(\ref{eq:injA}) nuclei. The exponential term in Eq.~(\ref{eq:nA-solut}) 
represents the survival probability during the propagation time $t'-t$ 
for a nucleus with fixed $A$ and can be computed according to 
Aloisio et al. 2012a. 
The derivative term $d\Gamma'/d\Gamma$ present in both solutions 
(\ref{eq:np-solut}) and (\ref{eq:nA-solut}) is analytically given (Aloisio et al. 2012a).

\section{Monte Carlo} 

The kinetic approach outlined above neglects interactions fluctuations considering an (average) continuum loss 
of energy suffered by particles. This approximation in the case 
of protons, has a limited effect on the flux computation only at 
the highest energies ($E>100$ EeV) (Berezinsky et al. 2006a, Berezinsky et al. 2006b, Aloisio et al. 2007).

\begin{myfigure}
\centerline{\resizebox{80mm}{!}{\includegraphics{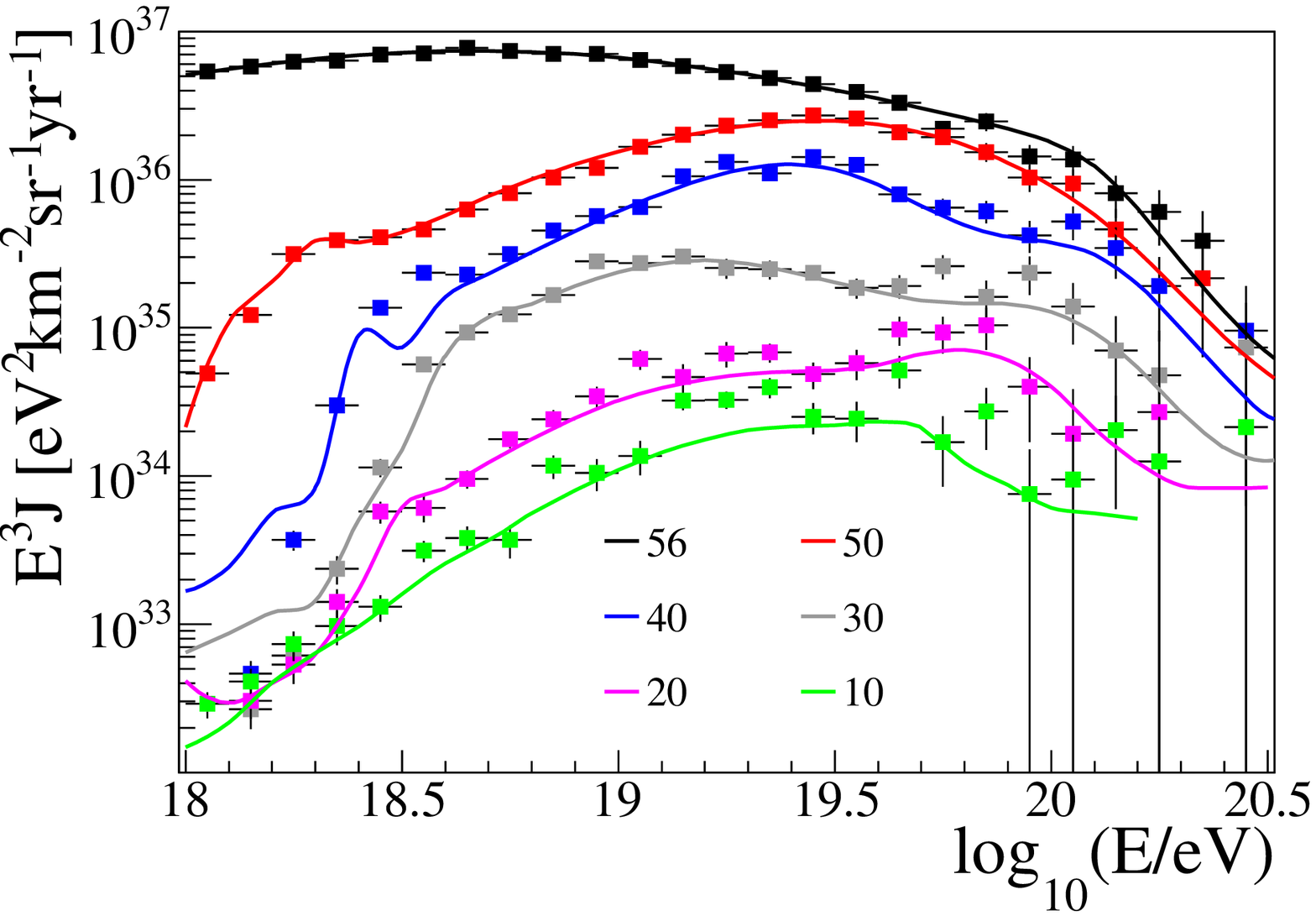}}}
\caption{\small{Flux of iron and secondary nuclei (A=50, 40, 30, 20, 10) at $z=0$ in the case of pure iron injection at 
the source with a power law injection index $\gamma=2.2$. Full squares correspond to the {\it SimProp} result 
(Aloisio et al. 2012b) while continuous lines correspond to the solution of the nuclei kinetic equation of Aloisio et al. 2012a.}}
\label{fig1}
\end{myfigure}

In order to evaluate the effects of fluctuations on the expected nuclei flux we have build a 
computation scheme alternative to the kinetic one, that uses the MC technique to simulate nuclei interactions.
First of all, let us remark that fluctuations could be relevant 
only in the case of nuclei photo-disintegration, this follows from the fact that the pair-production process 
involving nuclei can be considered as an interaction process of the inside nucleon, 
therefore being fluctuations in the protons pair-production irrelevant (Berezinsky et al. 2006b) the same holds for nuclei.  
The MC simulation scheme we have developed {\it SimProp} (Aloisio et al. 2012b) is mono-dimensional: 
it does not take into account spatial distributions tagging sources only through their distance from the observer
(red-shift). The MC simulation propagates 
particles in steps of red-shift following the injected nucleus, secondary nuclei and protons produced at 
each photo-disintegration interaction and calculating their losses up to the observer, placed at red shift zero.
The nuclear model on which {\it SimProp} is based is the same used for the kinetic approach 
(see Aloisio et al. 2012a, 2012b and references therein). The stochastic nature of the nuclei photo-disintegration 
process is modeled through the survival probability of a nucleus of atomic mass number $A$ and Lorentz factor 
$\Gamma$
\begin{equation}
P(\Gamma,z) = \exp \left(- \int^{z^{*}}_{z} \frac{1}{\tau_{A}(\Gamma,z^{'})}
\left|\frac{dt}{dz^{'}}\right| dz^{'} \right)  
\label{eq:prob} 
\end{equation}
where $z$ and $z^*$ are the values of the redshift of the current step (from $z^{*}$ to $z$).

The {\it SimProp} code is designed in such a way that any red-shift distribution of sources and any injection spectrum can be 
simulated. This is achieved drawing events from a flat distribution in the red-shift of the sources and of the 
logarithm of the injection energy. Once the event is recorded at z = 0 the actual source/energy distribution is recovered 
through a proper weight attributed to the event (Aloisio et al. 2012b). 
We will compare now the spectra obtained using {\it SimProp} (Aloisio et al. 2012b) with those calculated solving the kinetic 
equation associated to the propagation of nuclei (Aloisio et al. 2012a). To pursue such comparison, a pure 
iron injection with a power law injection of the type $\propto E_g^{-\gamma}$ with $\gamma=2.2$ have been assumed. 
The sources have been assumed to be homogeneously distributed in the red-shift range $0<z<3$. In figure \ref{fig1} the 
fluxes expected at $z=0$ are shown for iron and secondary nuclei produced in the photo-disintegration chain suffered
by primary injected irons. The points refer to the {\it SimProp} results while the continuous lines to the fluxes computed in 
the kinetic approach. A good agreement between the two schemes is clearly visible in figure \ref{fig1}. 
At the highest energies the path-length of iron nuclei is very short. Therefore, to achieve a good sampling in the MC 
simulation,  higher statistics is needed; this is the reason for larger errors bars
in the {\it SimProp} results at the highest energies. 
Let us conclude discussing why it is useful to go beyond the kinetic approach. The kinetic approach 
has the important feature of being analytical: fluxes are computed mathematically solving a first principles equation 
(Aloisio et al. 2012a). This means that the flux of primaries and secondaries is expressed in terms of several integrals that 
can be computed numerically, once the injection spectrum and the sources distribution are specified. In particular, the 
flux of secondary nuclei and nucleons produced in the photo-disintegration chain of a primary $A_0$ is obtained by the 
numerical computation of $A_0$ nested integrals and this computation should be repeated each time the hypothesis 
on sources (injection and distribution) are changed. This computation, while it is always feasible numerically, takes some 
time that can be substantially reduced using a MC computation scheme. This follows by the fact that, as discussed above, 
within the {\it SimProp} approach it is possible to simulate different source distributions and injection spectra without repeating 
the overall propagation of particles. In this sense the MC approach presented here, which is the minimal stochastic extension 
of the kinetic approach, provides a faster computation scheme. 

\thanks
I'm grateful to all collaborators with whom the works presented here were developed: V. Berezinsky, 
A. Gazizov and S. Grigorieva, from the theory group of the Gran Sasso Laboratory, D. Boncioli, A. Grillo, 
S. Petrera and F. Salamida, Auger group of L'Aquila University, P. Blasi, from the High Energy Astrophysics group of 
the Arcetri Astrophysical Observatory.

\bigskip
\bigskip
\noindent {\bf DISCUSSION}

\bigskip
\noindent {\bf CARLO GUSTAVINO:} The difference between Auger and the other experiments 
can be due to the fact they are looking from different hemispheres?

\bigskip
\noindent {\bf ROBERTO ALOISIO} This is an hypothesis that was recently put forward. I personally 
do not believe in such explanation because of the simple reason that at energies around $2\div 3\times 
10^{19}$ eV, where already the difference between Auger and HiRes starts, the universe visible in 
UHECR  has a huge scale of the order of Gpc. Therefore it is very unlikely to have differences between 
observations carried out from the southern and northern hemispheres.

\end{multicols}
\end{document}